\documentclass[12pt]{iopart}

\usepackage{bm}
\usepackage{graphicx}
\usepackage{color}
\newcommand{\xxx}{\vspace*{8pt}}

\begin{document}

\title[Prepint: How far can Tarzan jump?]
{How far can Tarzan jump?}

\author{Hiroyuki Shima}

\address{Department of Environmental Sciences and 
Interdisciplinary Graduate School of Medicine and Engineering, 
University of Yamanashi, 4-4-37, Takeda, Kofu, Yamanashi 400-8510, Japan}
\ead{hshima@yamanashi.ac.jp}

\begin{abstract}
The tree-based rope swing is a popular recreation facility, 
often installed in outdoor areas, giving pleasure to thrill-seekers.
In the setting, one drops down from a high platform, 
hanging from a rope, then swings at a great speed like ``Tarzan",
and finally jumps ahead to land on the ground.
The question now arises: How far can Tarzan jump by the swing?
In this article, I present an introductory analysis of 
the Tarzan swing mechanics, a big pendulum-like swing 
with Tarzan himself attached as weight. 
The analysis enables determination of how farther forward 
Tarzan can jump using a given swing apparatus. 
The discussion is based on elementary mechanics and, 
therefore, expected to provide rich opportunities for 
 investigations using analytic and numerical methods. 
\end{abstract}

\submitto{\EJP}

\section{Introduction}

An oft-repeated scene in Hollywood movies and adventure anime is 
that of someone jumping on to the end of a long rope suspended 
from above and, in pendulum motion, 
gallantly leaping over the various dangers below 
(enemies, wild animals, poisonous swamps, etc.). 
Very often, a similar children's play facility called 
the Tarzan swing \cite{Edwards1991} is set up in forest parks and beaches. 
Holding onto the end of a rope and jumping down from a high altitude is 
a moment that would test anyone's courage.

In terms of mechanics, the Tarzan swing can be defined as follows:
{\it ``It refers to a series of movements, 
starting from the forward jump to the actual landing, 
by using a half cycle of the pendulum's full swing with one's 
own bodyweight being the pendulum's weight".}
After reading this sentence,
a physics student may pose the following question:
How far can one travel in the horizontal direction with this acrobatic jump?
The deciding factor here is the moment where Tarzan releases 
the vigorously swinging rope. 
It would be intuitive that Tarzan can travel the maximum distance in the horizontal direction 
by releasing the rope at the precise right time (neither too early, nor too late).

This argument brings us to the main theme of this thesis. 
Let us recall that the rope hangs from a fulcrum and swings forward.
The angle formed between the swinging rope 
and the vertical line is expressed by a variable $\theta$.
Then, at which point of the pendulum's deflection angle $\theta$
should Tarzan take his hands off to travel the maximum distance 
in the horizontal direction?

If this problem is posed during a mechanics lecture, 
the students may respond as follows:
``When $\theta=\pi/4$, the maximum distance travelled can be attained".
This answer is correct in the case of, for instance,
firing a cannon from the earth's surface. 
According to the simple ballistic model, 
a cannon fired at a launch angle of $\pi/4$ 
from the ground would travel the greatest distance (Refer to Appendix). 
Even in the case of throwing a baseball, the ball would travel 
the greatest distance if thrown at an upward oblique angle of $\pi/4$,
as anyone would probably know from experience. 
However, this answer is incorrect in the case of a Tarzan swing. 
As shown below, for Tarzan to travel the maximum distance, 
it is essential for him to release the rope at 
a much smaller deflection angle than $\pi/4$. 
And interestingly, this optimum angle of deflection depends on 
the length of the rope used and the height of the platform 
({\it i.e.} the starting position). 
To arrive at this conclusion, a knowledge of elementary mechanics 
and some knowledge of numerical calculation is sufficient. 
Furthermore, through a simple experiment \cite{Trout2001} using a pendulum, 
it is possible to compare the theory with the data measured 
and study the physical cause of disparity between them. 
Therefore, it can be said that the above problem is 
a suitable research topic that can be assigned to 
students in senior high school or in the first year of college
to test their application of basic mechanics.\footnote{By the way, 
a different research theme is recommended for 
studying the case of Tarzan not releasing the rope 
due to fear or cowardice.
Actually, since ``swing mechanics" is extremely useful 
in a first year physics course,
there have been many publications of extremely interesting 
topics in this regard \cite{Piccoli2005,Post2007,Roura2010}.}

\section{Problem establishment}

\begin{figure}[bbb]
\centerline{
\includegraphics[width=10.0cm]{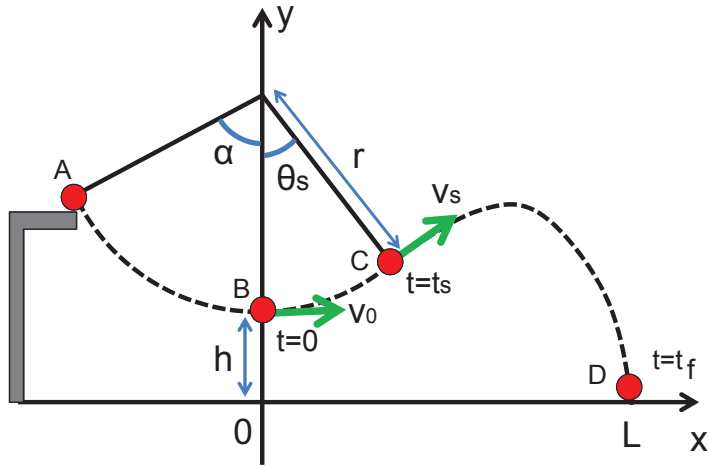}
}
\caption{(color online) Schematic view of the Tarzan swing motion.
Tarzan initially stands on the platform (A). 
He swings from this point and reaches the swing nadir (B), 
from where he swings ahead to the launching point (C), 
where he releases the rope to achieve as long a flight distance
as possible to reach the landing site (D).
All the variables and constants used in the present article
are displayed.}
\label{fig_system}
\end{figure}

For simplicity in the following discussion, 
consider Tarzan himself (who holds the rope) 
to be the weight with a mass of $m$.
Assume that Tarzan moves in the vertical plane 
described by the $x$-$y$ axis and that air resistance 
is negligible due to his sufficiently large mass.
Also we take no account of mass and deflection of the rope suspended.
Figure \ref{fig_system} shows: (i) a schematic diagram of 
the sequence of processes in the Tarzan swing from the take-off 
at a high platform until the landing, 
and (ii) the definitions of constants and variables used in this article. 
First, Tarzan jumps off the platform (Point A of Figure \ref{fig_system}), 
reaches the pendulum swing nadir B ($y = h$) at time $t = 0$ 
and releases the rope (Point C) at time $t = t_s$. 
We assume the pendulum's deflection angle to be $\theta = \theta_s$
and Tarzan's velocity to be $v = v_s$ at the moment he releases the rope.
Then his position at $t \ge t_s$, 
after he throws himself into the air, can be represented by the following equation.
\begin{eqnarray}
x(t) &=& r \sin\theta_s + v_s t \cos\theta_s, 
\label{eq_001} \\
y(t) &=& h + r (1-\cos\theta_s) + v_s t \sin\theta_s - \frac{g}{2}t^2.
\label{eq_002}
\end{eqnarray}
Here, $g$ is the acceleration due to gravity 
and $r$ is the length of the rope. 
After diving through the air, Tarzan finally lands on 
the ground at $t = t_f$. (Point D)

Based on the above conditions, the problem that we need to solve 
can be summarized as follows:
``Using the given constants, $h$, $r$, and $\alpha$, 
determine the swing deflection angle $\theta$ 
that maximizes the horizontal flight distance $L$,
and evaluate the maximum distance".

\section{Formulation}

Let us address the problem now.
At $t=t_f$, the Tarzan lands on the ground,
and thus, we have $x(t_f)=L$ and $y(t_f)=0$.
Substituting these equations into Eqs.~(\ref{eq_001}) and (\ref{eq_002}), 
we obtain
\begin{eqnarray}
L &=& r \sin\theta_s + v_s t_f \cos\theta_s, 
\label{eq_003} \\
0 &=& h + r (1-\cos\theta_s) + v_s t_f \sin\theta_s - \frac{g}{2}t_f^2.
\label{eq_004} 
\end{eqnarray}
We eliminate $t_f$ from Eqs.~(\ref{eq_003}) and (\ref{eq_004})
to obtain the expression of the flight distance $L$ as
\begin{eqnarray}
L &=& 
r \sin\theta_s + \frac{v_s^2 \sin\theta_s \cos\theta_s}{g} \nonumber \\
& &+
\sqrt{
\left( \frac{v_s^2 \sin\theta_s \cos\theta_s}{g}\right)^2
\!\!\!+\! \frac{2v_s^2 \cos^2\theta_s[h \!+\! r(1 \!-\! \cos\theta_s)] }{g}
}.
\label{eq_005}
\end{eqnarray}
Note that the launch velocity $v_s$ involved in Eq.~(\ref{eq_005})
is dependent on $\theta_s$.
The $\theta_s$-dependence of $v_s$ originates from
the energy conservation law represented by
\begin{equation}
mg [h+r(1-\cos\alpha)] = mg [h+r(1-\cos\theta_s)] + \frac{m}{2}v_s^2.
\end{equation}
It is reduced to a simpler expression:
\begin{equation}
\frac{v_s^2}{g} = 2r(\cos\theta_s-\cos\alpha),
\label{eq_010}
\end{equation}
which clarifies the relationship between 
$\theta_s$ and $v_s$.
The constraint given in Eq.(\ref{eq_010})
is the main reason that the optimal $\theta_s$ in the Tarzan swing
becomes smaller than $\pi/4$.
In the shell firing case, in contrast, $\theta_s$ and $v_s$
can be defined separately, as a result of which the optimal $\theta_s$
equals $\pi/4$ (see Appendix A).

\xxx

From Eqs.~(\ref{eq_005}) and (\ref{eq_010}),
we attain the expression:
\begin{equation}
\frac{L}{r} 
= \sin\theta_s + \Delta_2 \sin 2\theta_s  
+2 \cos\theta_s
\sqrt{
\Delta_2 \left(\Delta_2 \sin^2\theta_s+ \Delta_1\right)
},
\label{eq_018}
\end{equation}
where
\begin{equation}
\Delta_1 = \frac{h}{r} + (1 \!-\! \cos\theta_s), \quad 
\Delta_2 = \cos\theta_s-\cos\alpha.
\label{eq_020}
\end{equation}
In terms of physical meaning, $\Delta_1$ quantifies
the vertical height of the launching point 
({\it i.e.} point C in Fig.~\ref{fig_system}) in units of $r$.
$\Delta_2$ characterizes the difference in height
between the launching point and the starting point
(point A in Fig.~\ref{fig_system}).

\begin{figure}[ttt]
\includegraphics[width=7.7cm]{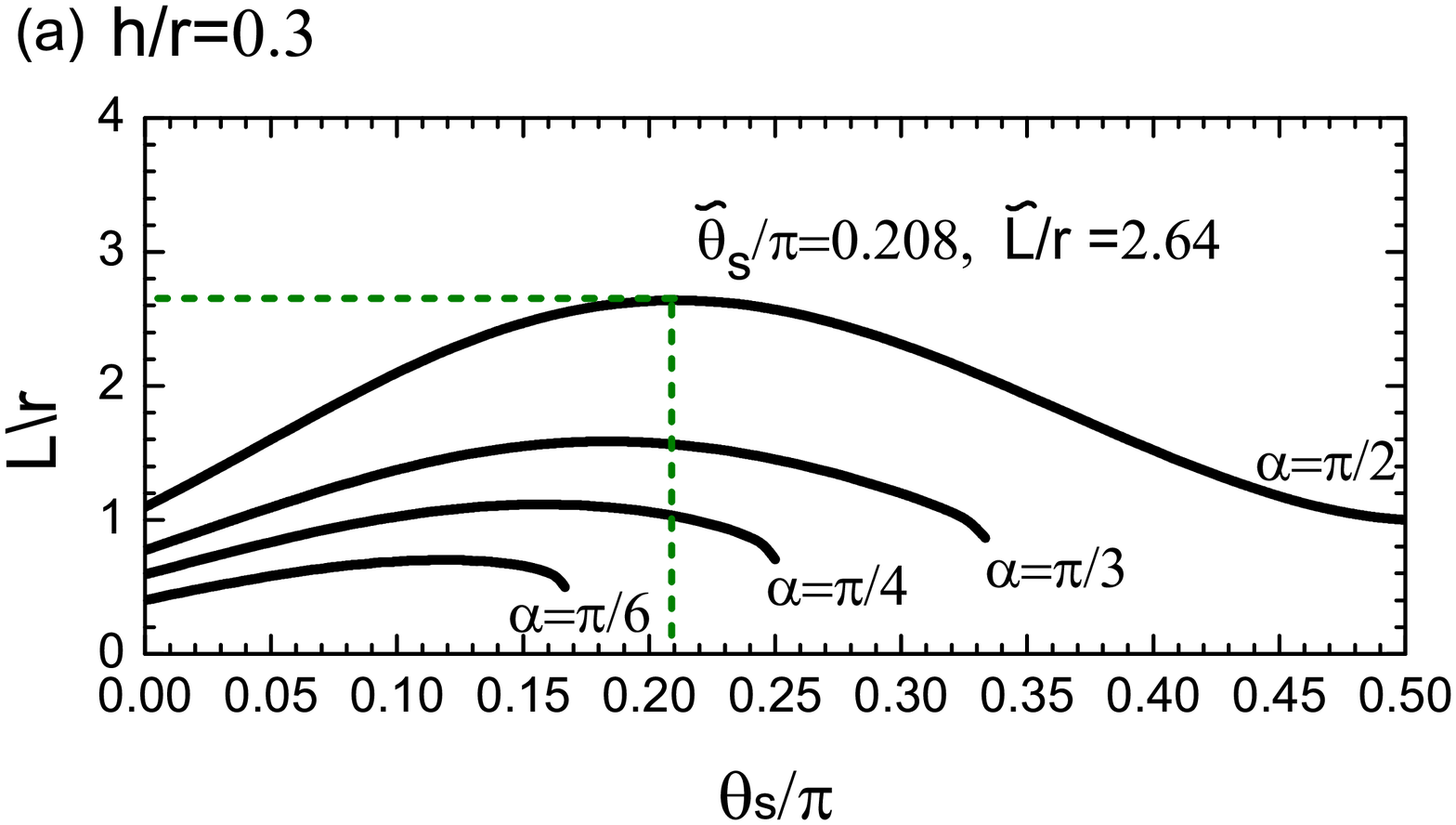}
\includegraphics[width=7.7cm]{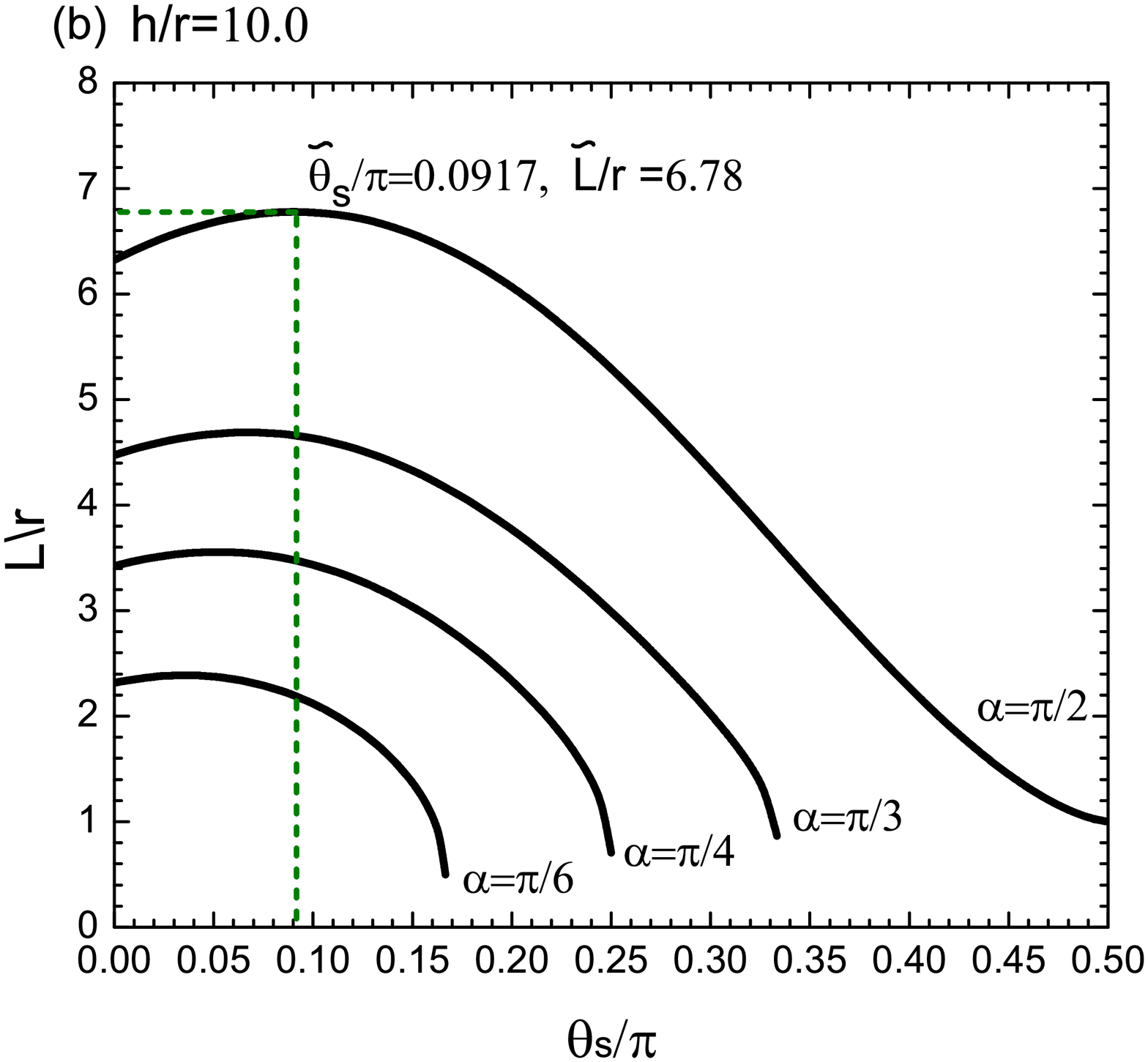}
\caption{(color online) Horizontal jump distance $L$
as a function of the launching angle $\theta_s$.
The nadir height $h$ in units of $r$ is set to be 
$h/r=0.3$ for (a) and $h/r = 10.0$ for (b).
The optimal values of $\tilde{\theta}_s$ and $\tilde{L}$
for the case of $\alpha=\pi/2$ are expressed in digits as examples.}
\label{fig_l_vs_theta}
\end{figure}

\xxx

Equation (\ref{eq_018}) provides the flight distance $L$
covered by Tarzan for a given $\theta_s$.
It is easy to derive from Eq.~(\ref{eq_018}) that
\begin{equation}
\frac{L}{r} = \sin\alpha 
\;\; \mbox{at $\theta_s=\alpha$,} \quad \mbox{and} \quad
\frac{L}{r} = 2\sqrt{\frac{h}{r}(1-\cos\alpha)}
\;\; \mbox{at $\theta_s=0$}.
\label{eq_025}
\end{equation}
In addition, we can prove that 
\begin{equation}
\alpha \left.\frac{dL}{d\theta_s}\right|_{\theta_s=0} 
> L(\theta_s=\alpha) \quad \left( 0<\alpha<\frac{\pi}{2} \right).
\label{eq_027}
\end{equation}
The inequality (\ref{eq_027}) guarantees the existence of at least one local maximum of $L$
within the range of $0<\theta_s<\alpha$;
the proof is left to readers as an exercise.
Henceforth, we use $\tilde{\theta}_s$
to denote the optimal launching angle that yields
the maximum flight distance $\tilde{L}$.

\section{Results and Discussions}

\begin{figure}[ttt]
\includegraphics[width=7.8cm]{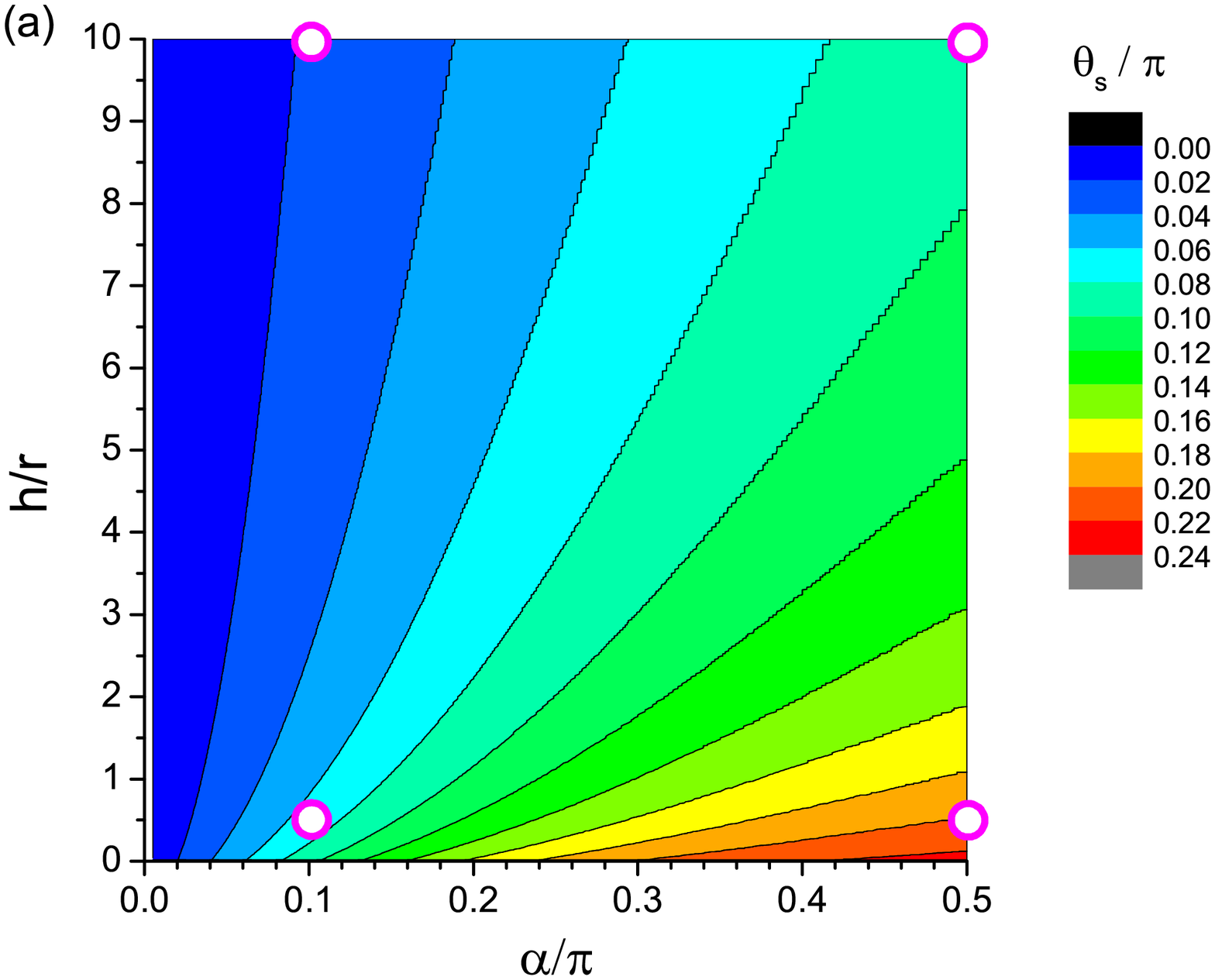}
\includegraphics[width=7.8cm]{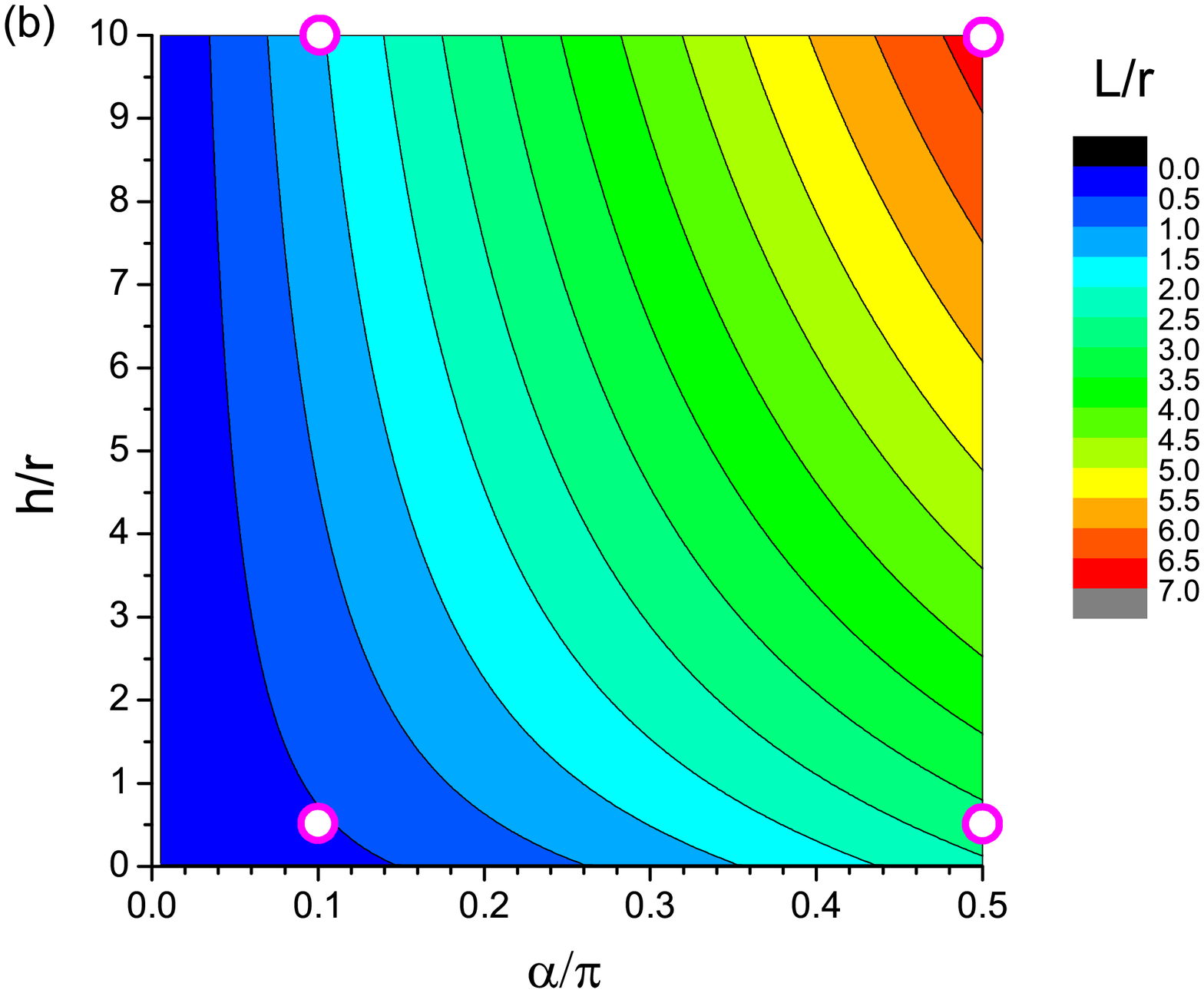}
\caption{(color online) (a) Contour plot of the $\theta_s$-surface
as a function of $h/r$ and $\alpha$.
(b) Contour plot of the $L$-surface.
Open circles depicted at the four corners indicate
the parameter conditions of the four trajectories 
shown in Fig.~\ref{fig_trajectory}.}
\label{fig_max}
\end{figure}

Figure \ref{fig_l_vs_theta} illustrates the $\theta_s$-dependence
of $L$ described by Eq.~(\ref{eq_018}).
Among the three parameters $\{h,r,\alpha\}$, 
the first two are set to have the relation $h/r=0.3$ in Fig.~\ref{fig_max}(a) and 
$h/r=10.0$ in Fig.~\ref{fig_max}(b).
The value of $\alpha$ is tuned as presented in the plots.
We observe that all $L$-curves are convex upward,
possessing a maximum $\tilde{L}$ at $\tilde{\theta}_s$ less than $\pi/4$.
In the two plots, the values of 
$\tilde{\theta}_s$ and $\tilde{L}$ for $\alpha=\pi/2$ are 
displayed as examples.
As $\alpha$ reduces from $\pi/2$,
both $\tilde{\theta}_s$ and $\tilde{L}$ decrease in a monotonic manner.
These decreasing behaviours are universal regardless of the condition of $h/r$;
this fact can be visually confirmed by Fig.~\ref{fig_max},
wherein the contour plots of
$\tilde{\theta}_s$ and $\tilde{L}$ on the $(h/r)$-$\alpha$ plane are presented.
From a physical viewpoint, the result in Fig.~\ref{fig_max}(b) is rather trivial; 
the larger (smaller) $\alpha$ yields the longer (shorter) flight distance $L$,
because such $\alpha$ enhances (diminishes) the impetus of Tarzan at the launching point
in accord with the energy conservation law [see Eq.~(\ref{eq_002})].
Similarly, it is obvious that a larger (smaller) $h/r$ results in a larger (smaller) $L$,
as it leads to long (short) flight time duration in which Tarzan move forward
in the horizontal direction.
On the other hand, the $\tilde{\theta}_s$-landscape shown in Fig.~\ref{fig_max}(a) 
is not so trivial and requires some consideration.
Why is the smaller $h/r$ associated with the larger $\tilde{\theta}_s$ and vice versa?
This question is resolved in part by depicting the trajectories
of Tarzan during the diving.

\xxx

\begin{figure}[ttt]
\centerline{
\includegraphics[width=13.6cm]{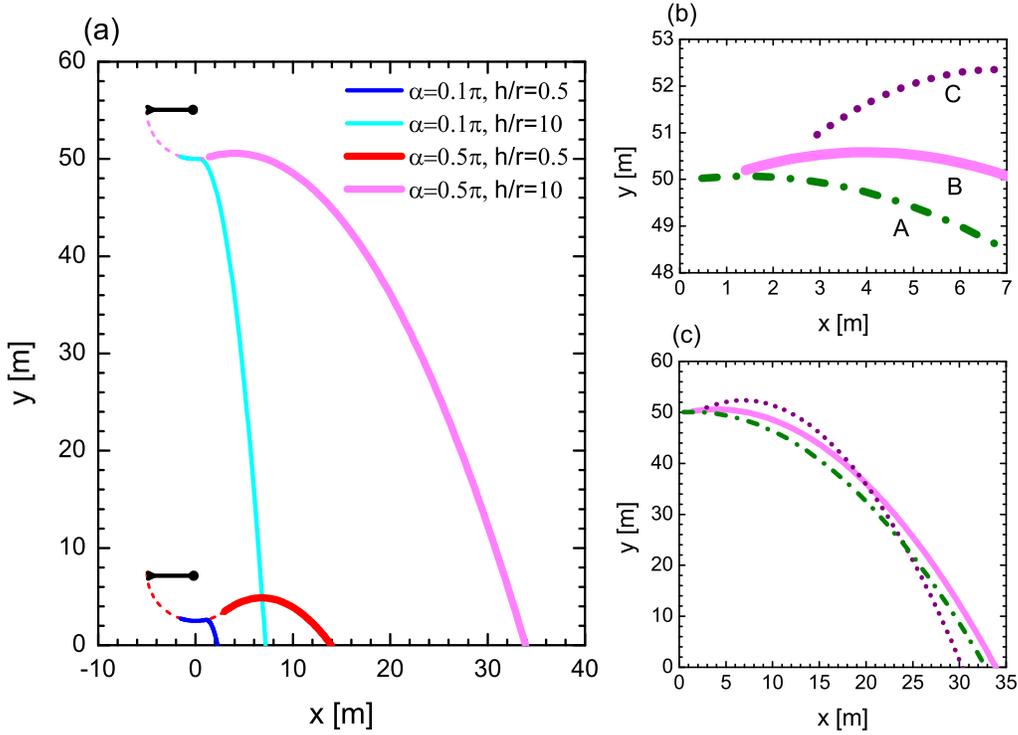}
}
\caption{(color online) (a) Trajectories of Tarzan under the four different
parameter conditions, labelled in Fig.~\ref{fig_max}.
Both axes are in units of meter, and $r=5$ m is selected for calculations.
(b-c) Three trajectories for $\alpha=\pi/2$ and $h/r=10.0$.
Each curve is associated to a different value of $\theta_s$.}
\label{fig_trajectory}
\end{figure}

Figure \ref{fig_trajectory}(a) shows the trajectories of Tarzan for four different 
conditions of the paired parameter ($\alpha$ and $h/r$), which correspond to the positions of
the four open circles depicted in Fig.~\ref{fig_max}.
The two horizontal arrows in Fig.~\ref{fig_trajectory}(a) symbolize
the ropes spanned by the pendulum pivot (circle) and Tarzan's hand
for the case of $\alpha=\pi/2$.
(A counterpart for $\alpha=\pi/10$ may be a slanted arrow with the same pivot,
but this is omitted in the figure.)
The figure makes it easy to grasp the two obvious facts
already found in Fig.~\ref{fig_max}(b).
First, for a fixed $\alpha$, the larger $h/r$ yields a larger $L$
owing to the long flight duration.
Second, when $h/r$ is fixed, a large $\alpha$ yields a large $L$
owing to  the large $x$-component of the launching velocity,
denoted by $v_s^x$.

\xxx

Moreover, Fig.~\ref{fig_trajectory}(a) provides a clue to the answer 
for the previously raised question, {\it i.e.,} the reason for
the smaller (larger) $h/r$ to correspond to a larger (smaller) $\tilde{\theta}_s$.
A small $h/r$ implies a short flight duration;
therefore, Tarzan prefers a large $\tilde{\theta}_s$ to obtain
a certain degree of the $v_s^y$ component in the upward direction,
because the upward $v_s^y$ component tends to prolong the flight duration.
This situation corresponds to the thick bottom curve in Fig.~\ref{fig_trajectory}(a), 
where Tarzan lifts off slightly upward like a flying ball to achieve $\tilde{L}$.
In contrast, when $h/r$ is large enough,
$\theta_s$-variation provides a minor contribution to the flight duration,
since Tarzan's vertical position $y(t)$ at $t\gg t_s$ is dominantly determined
by the gravitational force effect 
[{\it i.e.,} the term $-gt^2/2$ on the right side of Eq.~(\ref{eq_002})].
In other words, the flight duration is not significantly altered
by acquiring the upward $v_s^y$ component.
To reach maximum $L$, therefore,
it is crucial to obtain a large $v_s^x$ component by setting a small $\theta_s$.
However, one important aspect is to be kept in mind.
Even though a small $\theta_s$ is desired in principle,
a very small $\theta_s$ is not appropriate to maximize $L$.
This fact is schematically explained in Fig.~\ref{fig_trajectory}(b) and \ref{fig_trajectory}(c),
which demonstrate how the $\theta_s$-shift from the optimal value
affects the Tarzan's trajectory at $t\ge t_s$.
If Tarzan jumps at a very small $\theta_s$,
represented by the curve A, he necessarily lands before he successfully 
pulls out of the curve B ({\it i.e.}, the optimal case) in the horizontal direction.
Much worse is the case of a very large $\theta_s$ (curve C),
in which an insufficient $v_s^x$ component results in a short $L$
as compared with the other two cases.

\section{Summary}

We have developed a simplified model of the Tarzan swing.
The optimal deflection angle $\tilde{\theta}_s$ that maximize 
the horizontal jump distance of Tarzan was considered;
its dependency on configurational parameters of the swing apparatus
such as the height of the swing nadir ($h$), rope length ($r$), 
and starting platform position ($\alpha$)
was formulated in the realm of elementary mechanics.
We found that $\tilde{\theta}_s$ is a monotonic decreasing function of $h/r$;
the physical origin of the decreasing property
was explained through the trajectories of Tarzan during the aerial diving.

\section{Suggested problems}

It is reasonable to propose the following questions 
for further work related to this thesis.

\begin{itemize}
\item
In Fig.~\ref{fig_system}, consider a case where $h = r$, $\alpha=\pi/2$, and $\theta_s=0$. 
In this case, Tarzan travels through trajectories AB and BD, along each of which
he falls down at the same distance of $h$.
In addition, both the two initial velocities for the motions through AB and BD
({\it i.e.} velocities at points A and B, respectively)
have zero velocity component in the vertical direction ($v^y = 0$).
Now, determine the time required to travel from A to B (denoted by $t_{\rm AB}$) 
and B to D ($t_{\rm BD}$) and check whether they are equal. 
If they are not equal, give physical reasons for the inequality.
\item
In general, Tarzan at point B of Fig.~\ref{fig_system}
is subjected to large centripetal force (upward) due to speed $v_0$. 
What is the maximum value of the sum of gravitational downward-pull 
and centripetal force at point B? 
If the force applied is extremely large, then we fear the rope breaking or Tarzan slipping off.
\item
Carry out a study similar to this article for a case 
where the rope expands and contracts like rubber. 
Find out the values that can be set for the rope's elastic constants 
to maximize the horizontal distance covered by Tarzan.
\end{itemize}

\xxx

\section*{Acknowledgements}

The author acknowledges S.~Kitahara, C.~Hirata, A.~Kijima, K.~Yokoyama and Y.~Yamamoto
for private communications that sparked the author's interest in the present issue.
This study was financially supported by the inauguration allowance 
by University of Yamanashi.

\appendix

\section{The simplest ballistic curve}

We suppose a flying ball launched upward from the ground
at a slanted angle $\theta$ with respect to the horizontal line.
For a given initial velocity $v$,
the position of the ball at time $t$ is described by
\begin{equation}
x(t) = v t \cos\theta, \quad y(t)= v t \sin\theta - \frac{g}{2}t^2.
\end{equation}
At $t=t_f$, we obtain $x=L$ and $y=0$; thus we have
\begin{equation}
L = v t_f \cos\theta, \quad 0 = v t_f \sin\theta - \frac{g}{2}t_f^2,
\end{equation}
which imply that
\begin{equation}
L = \frac{2 v_s^2 \sin\theta_s \cos\theta_s}{g} = \frac{v_s^2 \sin 2\theta_s}{g}.
\end{equation}
Obviously, $L$ takes the maximum value at $\theta_s = \pi/4$.


\section*{References}

\end{document}